# The Relative Wavelength Independence of IR Time Lags in NGC4151 During the Years 2010-2015

V. L. Oknyansky[1]*, V. I. Shenavrin[1], N. V. Metlova[1], C. M. Gaskell[2]

[1] *Sternberg Astronomical Institute, M. V. Lomonsov State University, Moscow*

[2] *Department of Astronomy and Astrophysics, University of California, Santa Cruz*

**ABSTRACT**

We present results of a study of the correlation between the infrared (*JHK*L) and optical (*B*) fluxes of the nucleus of the Seyfert galaxy NGC 4151 for the years 2010 – 2015 using our own data (partially published) in combination with published data of Roberts and Rumstey (2012), Guo et al. (2014) and Schnülle et al. (2013, 2015). We find similar lags for each of the *HKL* passbands relative to the optical of 37 ± 3 days. The lags are the same to within the accuracy of measurement. We do not confirm a significant decrease in the lag for *HKL* in 2013-2014 previously reported by Schnülle et al. (2015), but we find that the lag of the short-lag component of *J* increased. We discuss our results within the framework of the standard model, where the variable infrared radiation is mainly due to the thermal re-emission of short-wave radiation by dust clouds close to a variable central source. There is also some contribution to the IR emission from the accretion disk, and this contribution increases with decreasing wavelength. The variability in *J* and *K* is not entirely simultaneous, which may be due to the differing contributions of the radiation from the accretion disk in these bands. The absence of strong wavelength-dependent changes in infrared lag across the *HKL* passbands can be explained by having the dust clouds during 2010-2015 be located beyond the sublimation radius. The relative wavelength independence of the infrared lags is also consistent with the hollow bi-conical outflow model of Oknyansky et al. (2015).

Key words: NGC 4151, IR and optical variability, cross-correlation analysis, dusty torus.

*oknyan@mail.ru

# INTRODUCTION

The nucleus of NGC 4151 is one of the most studied active galactic nuclei (AGNs), due to its brightness and significant variability at all frequencies, with the exception of radio frequencies. It has been intensively studied since the discovery of variability in 1967 (Fitch et al., 1967). The historical optical light curves from photographic plates dating back to 1906 (Oknyansky, 1978, 1983; Oknyansky et al., 2016; Pacholchik, 1971) and subsequent photoelectric observations (in particular observations by V. M. Lyutyi) are among the longest for any AGN. There are a number of IR and optical photoelectric observations dating back before 1967 (see references in Oknyansky 1993; Oknyansky and Lyuty, 2007; Oknyansky et al., 2016). The variability of NGC 4151 was discovered back in 1958 from photoelectric observations, but these results were published only 10 years later (de Vaucouleurs, de Vaucouleurs, 1968), when the variability of the object was already established. NGC 4151 is in the original list of "Seyfert galaxies" published by Seyfert (1943), and the first spectral observations were made 100 years ago (Campbell and Moore, 1918). Cherepashchuk and Lyuty (1973) discovered the variability of the broad H$\alpha$ line in NGC 4151, and also made the first measurement of time lags in the variability of AGNs.

Since the discovery of quasars in 1963, astronomers have tried to introduce some sort of order into the variety of AGN types. Khachikyan and Weedman (1971) divided Seyfert galaxies into two types: type 1 being objects with very broad permitted emission lines with full widths at half maximum (FWHM) typically of ~ 2000 to ~ 10,000 km / s) and with relatively narrow forbidden emission lines with half-widths of up to 2000 km/s; and type-2 objects for which the spectrum shows only relatively narrow emission lines. The nucleus of NGC 4151 was a typical type-1 Seyfert (Sy1) and the nucleus of NGC 1068 nucleus is a typical type-2 Seyfert (Sy2). A major breakthrough in understanding the differences between type-1 and type-2 AGN occurred in the 1980s following the discovery by Keel (1980) that the appearance of an AGN's optical spectrum correlates with orientation. This led to the unified model of AGN, which in a simplest form postulates that type-1 and type-2 s AGNs are actually the same type of object whose observational properties differ only because of the different orientation to the observer. This model, after the work of Antonucci (1993), is called the "straw person model" (SPM). The most important element of the SPM is the existence of optically-thick dust torus away from the poles of the AGN. It probably has a clumpy structure (Hönig, Kishimoto, 2011). The presence of such dust is a key factor for explaining the absorption of radiation in broad lines and the observed difference in the spectra of Seyfert nuclei of types 1 and 2.

Despite the significant success of the SPM, it also has problems. In particular, in 1984 observers found that the nucleus of the prototypical Seyfert 1, NGC 4151, had changed its type from Sy1 to Sy2 (Lyutyj et al., 1984; Penston and Peres.1984; Chuvaev and Oknyansky, 1989). It then returned, after a while, to being a Sy1 (Oknyansky et. al.,, 1991). At present, several dozen such cases of "changing-look" AGNs (CL AGNs) have been recognized, which allows us to conclude that this is not just a one-time phenomenon, but is relatively common. Obviously, the orientation of the object cannot change so quickly and thus CL AGNs are a serious problem for the simple orientation-unification model. The IR radiation of AGNs is dominated by thermal emission of dust heated by radiation from the inner regions of the accretion disc – emission that is energetically dominated by the extreme UV. The shape and structure of the "torus" is uncertain. Although it is commonly depicted in cartoons as being like a doughnut. Spatially-resolution IR observations of a number of AGNs show that dust clouds predominantly emitting in the mid-IR and far-IR range are not concentrated in the plane of the galaxy or of the accretion disc, as expected, but in polar regions (Braatz et al., 1993; Cameron et al., 1993; Bock et al., 2000; Hönig et al., 2012.

The main method for studying the unresolved structure of the emission from warm dust is reverberation mapping using IR and optical (UV) variability data. NGC 4151 was the first AGN for which such a lag was assumed on the basis of a visual analysis of light curves (Penston et al., 1971). This lag was interpreted as a consequence of the spatial remoteness of the dust heated by the variable radiation from the central source. The lag was later measured using the standard cross-correlation analysis of series of observations (Oknyansky, 1993; Oknyansky et al., 1999). The first measurement of the lag in variability in the $K$ band with respect to the optical in NGC 4151 gave a lag of 18 days (Oknyansky, 1993). The lag was longer, ∼ 26 days, for the $L$ band from the same data (Oknyansky and Horn, 2001).

The IR lag for NGC4151 varies with the level of luminosity of the central source, but with a delay of some years (Oknyansky et al., 2008, Kishimoto et al., 2013). The change in the lag can occur as a result of sublimation and dust recovery processes with changes in the level of the UV radiation of the nucleus. At present, such studies have been carried out for several dozen AGNs. Of great interest is the study of the dependence of the magnitude of the lag on the wavelength in the infrared range. For NGC 4151, in our first IR reverberation papers (Oknyansky et al., 1999; Oknyansky, 2002, Oknyansky and Horn, 2001) it was noted that the lag in the longer-wavelength $L$ band was significantly greater than the lag in the shorter wavelength $K$ band. After the outburst of the nucleus in 1996, the IR lag significantly increased and, interestingly, the lags in the $K$ and $L$ bands were identical to within the limits of measurement accuracy (Oknyansky et al., 2006). In our more recent papers (Oknyansky et al., 2014a, b) we again found that the lags in all IR bands of *JHKL* were the same to within the accuracy of the measurements. Our analysis of the published photometric data for a number of other AGNs has shown that this relative independence of lags with IR wavelength is more the rule than the exception (Oknyansky et al., 2014a, b). We suggested that a significant increase in IR lag with wavelength can be observed during a period of significant growth of the luminosity of the central source, when dust sublimation occurs and the lag value increases. Because of the delay of several years in changes of the size of the lags, an increase in the magnitude of infrared lags with a wavelength can be observed after a few years after a major outburst. A straight-forward explanation would be that the dust clouds are located beyond the region of possible dust sublimation at the current luminosity level. Obviously, for most objects this situation is realized, when a major outburst occurred in the past and not in the time interval being studied.

The possibility of determining cosmological constants based on the lag of infrared variability was first mentioned by Kobayashi et al (1998) and was independently proposed and first implemented by Oknyansky (1999, 2002). In recent publications of Yoshii et al. (2014), Hönig (2014), Hönig et. al. (2014) and Koshida et al. (2017), this method is considered in detail and practically applied (see also the discussion in Oknyansky et al., 2014b). Our discovery of the relative independence of the infrared lag from the wavelength for most AGNs is important in the practical application of this method, since it reduces the problem of a change in lag because of the shift in the rest-frame wavelengths of AGNs as a function of the redshift (for small $z <= \sim 0.2$). The application of this method to determine the distance to NGC4151 has shown that this AGN is located at a significantly larger distance than previously thought (Hönig et al., 2014).

The present paper is a continuation of our study of the lags of the IR passbands with respect to the optical in NGC 4151. We present results based on new monitoring in the *JHKL* bands combined with optical photoelectric and CCD photometry for the period 2010-2015. For a detailed review of the research and a description of the observational methods and data analysis the reader is referred to Oknyansky et al. (2014a). In addition to our own observations we make use of IR and optical data published of Schnülle et al. (2015). The combination of the data sets helps to make the light curves more complete and hence gives a more reliable determination of the infrared lags. Lags are determined for the time series using our MCCF code as discussed in previous papers in this series. We also use the JAVELIN - Monte Carlo Markov Chain method of Zu et al. (2013).

**OBSERVATIONAL DATA**

The method of IR observations in *JHKL* bands has been described in detail in our previous papers (see, for example, Taranova and Shenavrin, 2013). In the present study, we use our observations only for the 2010-2015 interval. Observations up to 2011 have been published in tabular form in Taranova and Shenavrin (2013), and observations up to 2015 are available on-line in an open access archive at http: http://www.sai.msu.ru/basa/inf.html. We have already analyzed data from 2008-2013 in an earlier papers (Oknyansky et al., 2014a, b). Similarly we add here new observations for 2013-2015, and also combined them with the published IR and optical data of Schnülle et al. (2013, 2015). These additional data significantly improved the light curves in the studied interval. Data for 2008-2009 were not included in this study due to their scarcity and lower accuracy. In addition, unlike previous publications, we did not include optical photometry data, obtained (mainly in 2009) with the 1.5-meter Maidanak telescope, because of the impossibility of reliably matching these data with our measurements (optical photometry of the object was not carried out by us during 2009). Most of the infrared observations have an accuracy of not worse than 1-2%. Compared to our earlier papers, the accuracy of our IR measurements improved during the interval being studied because of a number of upgrades of the equipment. In rare cases, the measurement errors were greater, but for further analysis we use only measurements with errors of no worse than 7%. In total, for the period 2010-2015, the average numbers of measurements per night in the *J*, *H*, *K* and *L* bands were 66, 54, 54, 66 respectively. Our data in JHK bands were supplemented by measurements of Schnülle et al. (2013, 2015) for 29 dates for 2010-2014. We reduced their data to our system. The method of optical photoelectric observations remained the same as that used by V.M. Lyutyi up to 2008 (see Oknyansky et al., 1999), but additional CCD measurements were used from a 60-cm telescope at the Southern Station of the SAI (see description in Oknyansky et al., 2012). Since there are no *U* band CCD measurements, we used the data in the band *B* to construct a composite optical light curve. The CCD observations were reduced to a system of photoelectric measurements with a 27" aperture. As in our previous papers (see Oknyansky et al., 2014a, b, 2015) *B*-band data were supplemented with published CCD *B*-band measurements of Roberts and Rumstey (2012) and 17 nights from Guo et al. (2014). In addition, we used 29 nights of optical measurements of Schnülle et al. (2013, 2015) in the red *z* band (29 dates) which were also reduced to our system B. Thus, all optical and infrared data are combined into one system. The formal accuracy of photoelectric measurements is generally not worse than 1-2%, but systematic differences between the measurements obtained on different instruments are possible. According to our estimates, these errors do not exceed 10% in the combined *B*-band light curve constructed for the 197 dates of the *B* value. The combined light curves NGC 4151 for 2010-2015 for *JHKL* and *B* are presented in Fig.1.

As it can be seen in Fig.1, variations in the brightness in the *JHKL* bands occurred almost synchronously, without any noticeable differences or shifts. There are fewer points in the *L* band, since Schnülle et al. (2013, 2015) have no observations in this band. In the changes there are rapid variations (with a characteristic time of tens of days) and a long-term trend (with a characteristic time of several years or more) with a maximum at the beginning of the time period. The optical light curve also exhibits rapid and slow changes with the same characteristic times, and the slow trend is more noticeable than in the IR light curves. In addition, rapid changes of a small amplitude with a characteristic time on the order of several days are noticeable in the optical variability.

During the entire time interval, the observed amplitude of variability in *B* was about 0.93 mag, whereas the variability amplitude from the historical light curves (Oknyansky et al., 2016), is almost 2 mag. At the same time, at the beginning of the time period being considered here, NGC 4151 was brighter than $12^{th}$ mag, which allows us to speak about the high activity of the object at this time. During the period 2010-2015, the brightness of NGC 4151 faded on average, but there was variability on timescales on the order of several months. We have not evaluated the contribution of underlying galaxy light in the 27" diaphragm in this paper, since this constant contribution does not matter for carrying out cross-correlation analyses. Also, estimates of the starlight contribution have been made previously, in particular in the work of Lyuti and Doroshenko (1999). If we take from their paper *B* = 12.91 for the host galaxy, then in the interval studied by us, the contribution of the AGN at the minimum values of $B = 12.81$ was about 10% of the recorded flux in $B$ and ~ $1/9^{th}$ of the contribution of the galaxy starlight. At a maximum brightness of about $B = 11.88$ mag, the contribution of the AGN was ~ 1.5 times brighter than the contribution of the galaxy. Thus, the flux of radiation from the central source changed by more than a factor of 13 in the $B$ band over the period 2010-2015. Heating of dust is mainly carried out by UV radiation (because the absorption cross section is higher), and we assumed that the variations of UV and optical radiation occur effectively synchronously (Edelson et al., 2017), and that the lag of optical variability relative to the UV (~ 0.5 days) is insignificant.

We plan to publish in a tabular form all the IR and optical observations used in the near future.

## METHODOLOGY OF CROSS-CORRELATION ANALYSIS

Cross-correlation analysis of astronomical time series presents difficulties because the time series are often unevenly sampled. To analyze the series, we applied our MCCF code, which is an upgrade of the method of Gaskell and Sparke (1986). The methodology of our analysis has not changed, and is described in detail in previous papers (see, for example, Oknyansky et al., 2014a, b, 2017). In the MCCF method, we strive to introduce a minimum number of arbitrary parameters, and also significantly reduce the contribution made by interpolation errors. A detailed discussion of other methods of cross-correlation analysis of non-uniform series, and their comparison with the MCCF method, was carried out in our previous papers (Oknyansky et al., 2014a, b, 2017).

We also carried out an analysis of the time series using the JAVELIN Monte Carlo Markov chain code of Zu et al. (2011, 2013). Further details and links can be found in Oknyansky et al. (2017). The JAVELIN method is not a very common method of cross-correlation analysis. In this method, the light curves are simulated 10,000 times on the basis of assumptions about the properties of the AGN variability which is assumed to be a damped random walk (see Gaskell and Peterson 1987), and then lags are found for each pair of these simulated light curves. A histogram is constructed of the lags thus found, which is used to find the optimal lag and its error (see, for example, Shappia et al., 2014; Zu et al., 2011, 2013). Of course, this method has similar problems as any method using interpolation and extrapolation of the time series. Nevertheless, this method has been used recently in some studies and it usually yields similar results as other methods. We decided to use this method as an additional check of our results given by the MCCF method, and without any changes in the authors' code. In the future, we plan to modernize JAVELIN in a manner similar to what was done in the MCCF method to reduce the contribution of interpolation errors in modeling the missing real observational data in the light curves.

**CROSS-CORRELATION OF LIGHT CURVES**

Cross-correlation functions of MCCF for the combined light curve in B and the variability in the *JHKL* bands in the period 2010-2015 are shown in Figs. 2 and 3. It can be clearly seen that the main maximum for all these cross-correlation functions is in the region of ~37 days. The magnitude of the lag of the variability at *K* relative to the optical remained almost the same as we found for the interval 2008-2013. Also, as in our earlier paper, the lag, within a measurement accuracy of ~ 3 days (see below) is practically independent of the wavelength. But at the same time, the form of cross-correlation functions in the region near zero lag is noticeably different. One can notice that the secondary maximum and its significance fall as one goes to increasingly longer wavelength bands from *J* to *L*. For *J*, this maximum corresponds to the lag about 4-6 days and is possibly associated with the variability of the accretion disc in the infrared. Fig. 4 shows the results of analysis using the JAVELIN method. From the comparison of Figs. 2, 3 and 4, it can be seen that both methods give approximately the same results.

The density of observations, unfortunately, is insufficient for analysis in shorter time intervals with the exception of the years 2013-2014, where optical and IR data are more homogeneous. MCCF cross-correlation functions for the combined light curves in *B* and *JHK* for these years are shown in Fig.5. The maxima in the cross-correlation functions for *K* and *H* remained approximately the same as in Fig. 2, but the correlation coefficients at the maxima became larger. In the cross-correlation between J and B, there are changes: the maximum at cross-correlation has shifted to the region of short lags. For the *L* band, calculations were performed in this interval, but are not shown. No significant changes are observed in comparison with Fig. 3. Independent analysis of the data for 2013-2014 using the JAVELIN method gave similar results (see Figure 6). Our analysis did not confirm the result of Schnülle et al. (2015) that there was decrease in the lag in this interval for the bands *K* and *H*, but for the band *J* the lag became noticeably smaller.

Fig. 7 shows light curves in *B* and *K* where the *K*-points were shifted to take into account the 38-day lag (corresponding formally to the maximum correlation) and linearly scaled to the *B*-band scale. Visual analysis shows good agreement of the variability in these two bands after shifting, and also shows that the lag determined does not depend on any one event in the light curves. Variations in *B* and *K* are in good agreement throughout the interval and in several characteristic periods of brightness changes.

**EVALUATION OF ERRORS**

To estimate the errors in the lags derived by the MCCF, we applied the same Monte Carlo procedure as before (Oknyansky et al., 1999, 2014a, b). Our estimates of the errors in the lags gave a value of about 3 days. Thus, the delays in the *HKL* bands with respect to the *B* band are $37 \pm 3$ days.

The histograms obtained in the JAVELIN method can be used to estimate the optimal lags (JAVELIN) and their mean-square errors (see details, e.g., Shappee et al., 2014; Zu et al., 2013)

**DISCUSSION**

This paper is a continuation of our series of studies on the correlation between infrared and optical variability in NGC 4151, as well as the variability of the IR lag values and their dependence on the wavelength (Oknyaskii, 1993; Oknyansky et al., 1999, 2006, 2008, 2014a, 2014b). In these studies we found that the delay in the *K* band is different in different activity states of the AGN, and also that the ratio of the lag values in the bands *L* and *K* varies considerably in the range 1-3. The results obtained in these studies were independently partially confirmed in other investigations (Koshida et al., 2009; Kishimoto et al., 2013; Schnülle et al., 2013; Hönig, Kishimoto, 2011).

Currently, there are IR and optical light curves of the object for decades. Using all of these data together for IR lag study would not lead to better statistics for several reasons. Firstly, the data are not homogeneous, secondly, during this time there were significant variations in the luminosity of the nucleus, thirdly, the lag values themselves and the shape of the response functions changed significantly. Our task, on the contrary, was to obtain estimates of delays in as short as possible time intervals.

A detailed historical review and discussion of the obtained results was carried out in a previous paper (Oknyansky et al., 2014a). In the present paper, using more complete observational material and using two independent methods of analysis, we confirmed the relative independence of infrared lags from the wavelength during 2010-2015. We do not confirm the significant decrease of the lags in 2013-2014, noted by Schnülle et al. (2015) for the *H* and *K* bands, but we found significant changes for the *J* band, which may be due to the relatively large contribution of the variability of accretion disk (AD) in this band with a strong dimming of the luminosity of the nucleus. The relative independence of the *HKL* lags from the wavelength can be related to the fact that the radius at which dust sublimation might occur is less than the distance to the dust clouds nearest to the center of the AGN. In fact, if the sublimation radius is $\sim L_{UV}^{-1/2}$, then, compared with the maximum in 1996, the luminosity in the UV during 2010-2015 was at least 10 times less (the amplitude of variability in UV is even greater than in the optical). Accordingly, if the lag in *K* was about 70-100 days (Oknyansky et al., 2008; Kishimoto et al., 2013) after the outburst in 1996, then the radius of sublimation was less than 20-30 light days in 2010-2015. This means that the dust clouds are located beyond the radius where sublimation might occur. This can explain the observed relative independence of the time delays with wavelength, and the absence of significant variability in the time delay values with changes in luminosity over a 10-year period (Oknyansky et al., 2014a, b; Schnülle et al., 2013; Hönig and Kishimoto, 2011). Thus, dust recovery after the outburst in 1996 led to a decrease in the lag value by more than a factor or two, but at distances closer than 30 light days the dust did not recover. Apparently, with strong bursts of the AGN, the dust was completely sublimated to a radius of 30-40 light days. In more remote areas, dust was partially survived in the denser interior parts of the clouds (Barvainis, 1992). Thus, in the clouds where the dust partially survived, it was more quickly restored when the luminosity falls. At closer distances, dust recovery requires more time, since it was completely sublimated there during the very bright and prolonged outburst in 1996.

The relative independence of infrared lags with the wavelength can also be explained in the model of bi-conic outflow of dust clouds that we proposed earlier (Oknyansky et al., 2015 ; Gaskell and Harrington, 2018). In this case, dust clouds emitting in the infrared range can have different dust temperatures at different distances from the central source, but the delays of the variability of the IR radiation from these clouds will be approximately the same. This interpretation can also be combined with the first explanation, that is, the dust clouds move from the center in the bi-conical regions, but are located further than the current radius of the sublimation zone. In this model, there is a natural explanation for the change in the spectral type of the nucleus in the formation of new dust in the clouds on the line of sight when UV radiation is weakened. Recent IR interferometric observations of the AGNs (Hönig et al., 2012, Hönig and Kishimoto. 2017; López-Gonzaga et al., 2016; Leftley et. al., 2018) confirm that the IR-emitting clouds are predominantly elongated along the polar regions, and are not located in the plane of the AD, where a gas-dust torus was presumed to be located. Another possibility to explain the relative independence of lags from the wavelength is the possibility that dust particles can have different temperatures depending on their size, and at the same time, approximately have the same localization. Similarly, in the Nenkov model (2008a, b), the temperature in gas-dust clouds can be inhomogeneous, and therefore each cloud emits IR radiation in a broad range, which can significantly weaken the dependence of infrared lags on the wavelength (see also the discussion of this problem in Vazquez et al., 2015; Hönig, Kishimoto, 2011; Pozo Nuñez et al., 2014).

In the last few years (after Oknyansky et al., 2014a, b), a number of studies have been carried out in which new results have been obtained on the relative independence of IR delays from the wavelength in the AGN, and various models for interpreting these results have been discussed (Oknyansky et al, 2015; Vazquez, 2015; Vazquez et al., 2015; Almeyda, 2017). The most detailed theoretical analysis involving various effects associated with geometries and the structure of the emission region, anisotropy of UV radiation, and cloud orientation was carried out by Almeyda et al. (2017). Taking into account the anisotropy of the AGN radiation field, the dust sublimation radius is a function of the polar angle and the region in which dust is absent is not a sphere, but rather resembles a figure eight in cross-section (see Fig.10 in Almeyda et al., 2017). In this case, the clouds may be closer to the source near the equatorial plane, in contrast to the situation of isotropic radiation from a central point source. Simulation of response functions at various IR wavelengths has shown that delays at different IR wavelengths in most cases do not change significantly with the wavelength.

We note that all these different interpretations of the relative independence of infrared delay from wavelength can be reconciled in one model, since the model of biconical outflows of dust clouds also assumes the presence of a torus, as well as the effects of radiation anisotropy, cloud orientation effects, and also the existence of a temperature dispersion of dust particles in a single cloud. However, the creation of a self-consistent model of IR radiation near the AGN is beyond the scope of this article. We note the need for further intensive research on the IR variability of AGNs that is of crucial importance for understanding not only the physics of these objects, but also cosmological applications.

We are grateful to S. Hönig for useful discussions, and also to D.-F. Guo for providing optical photometric data.

This work is devoted to the memory of Olga Taranova (1938-2017), our co-author in many studies of the IR variability of NGC4151 over the decades.

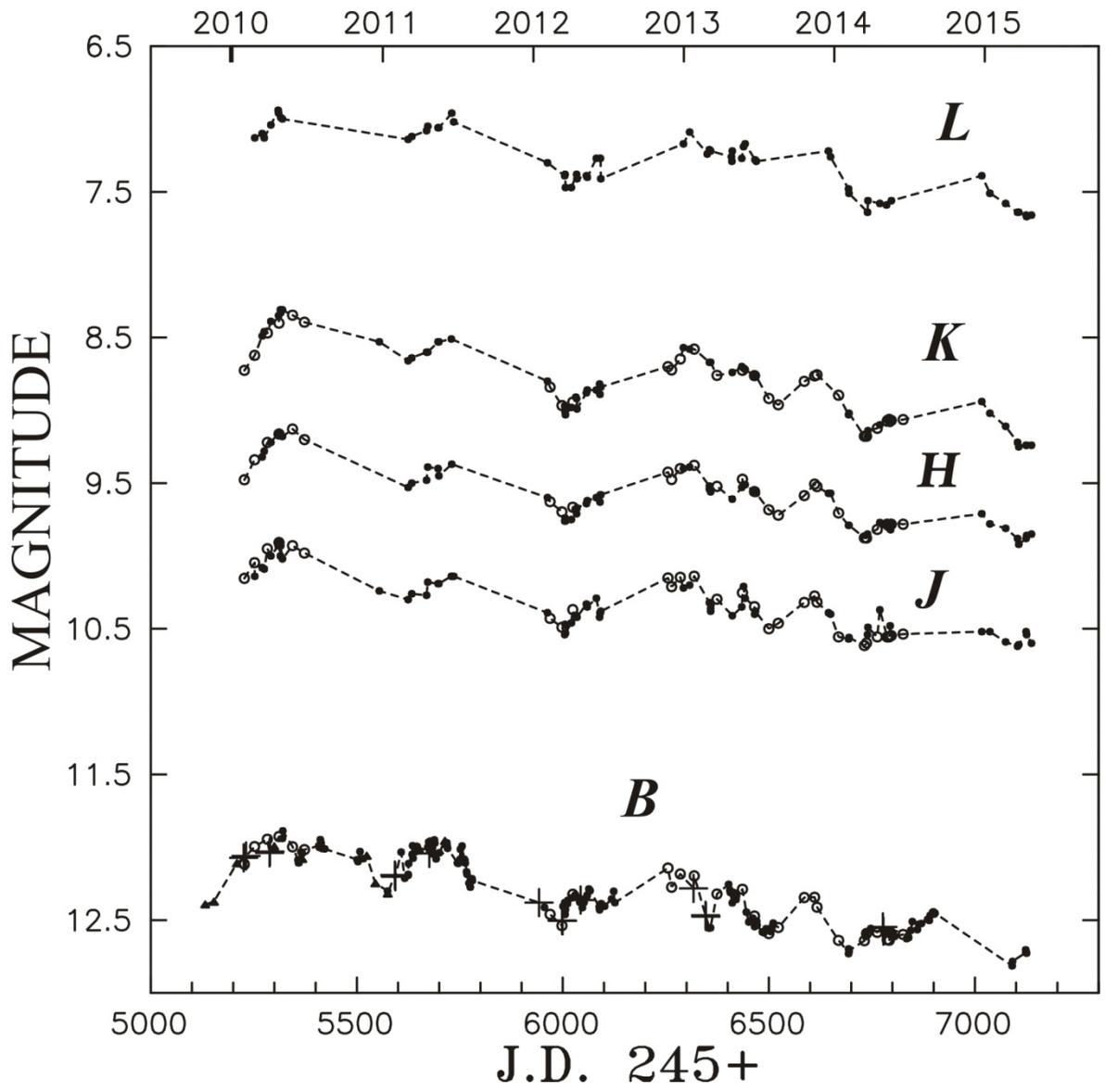

**Fig.1.**

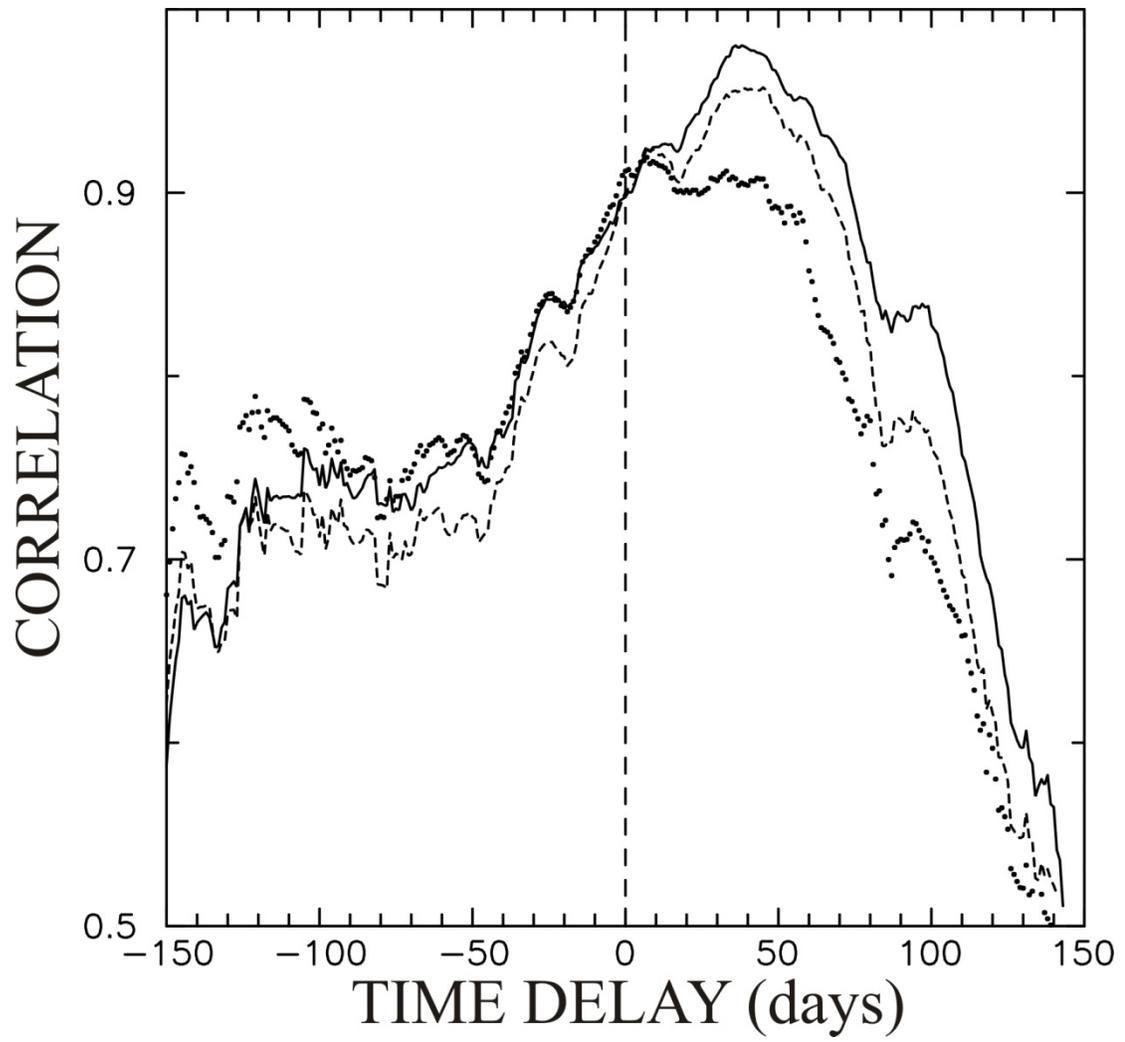

**Fig. 2.**

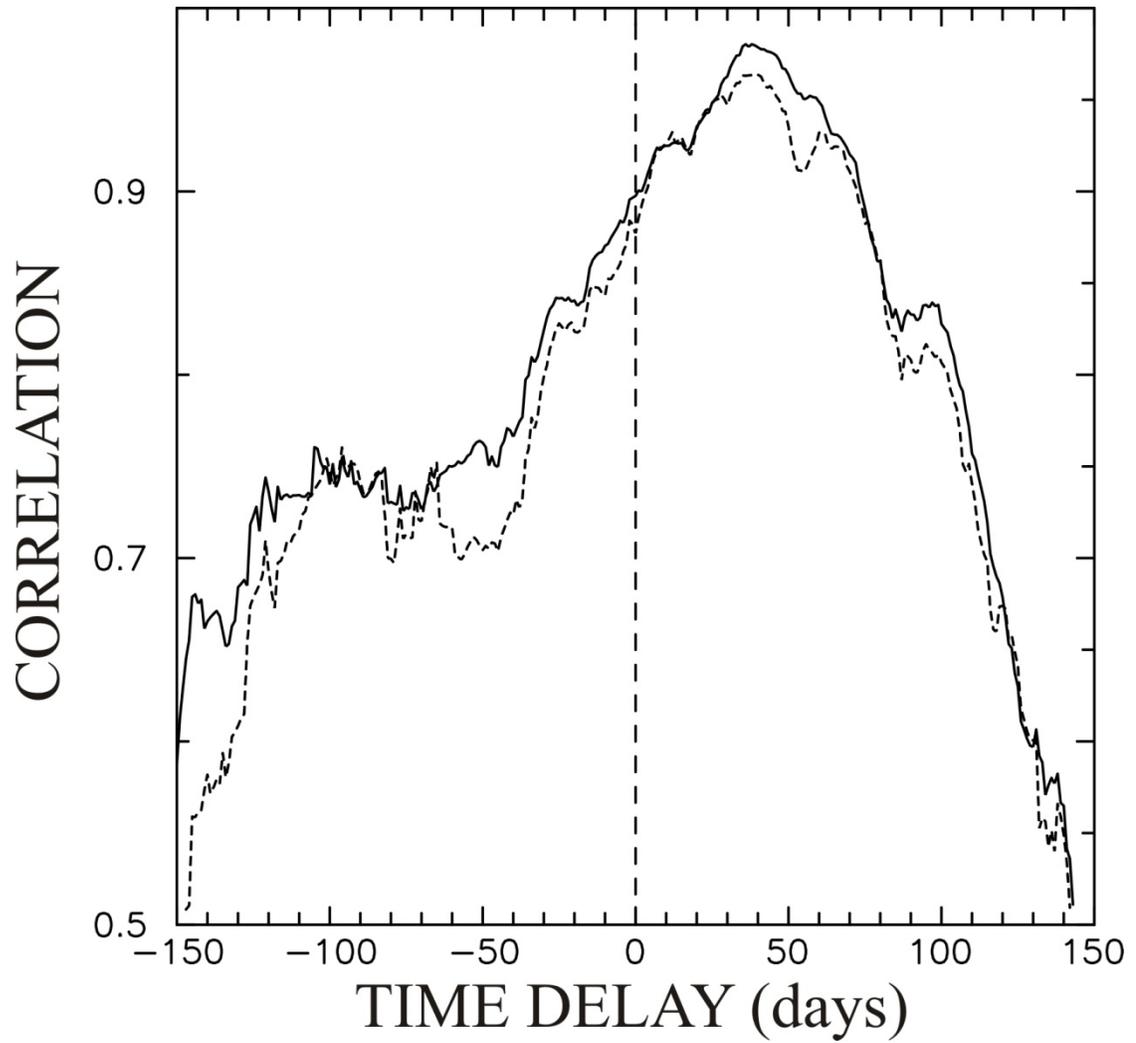

**Fig. 3.**

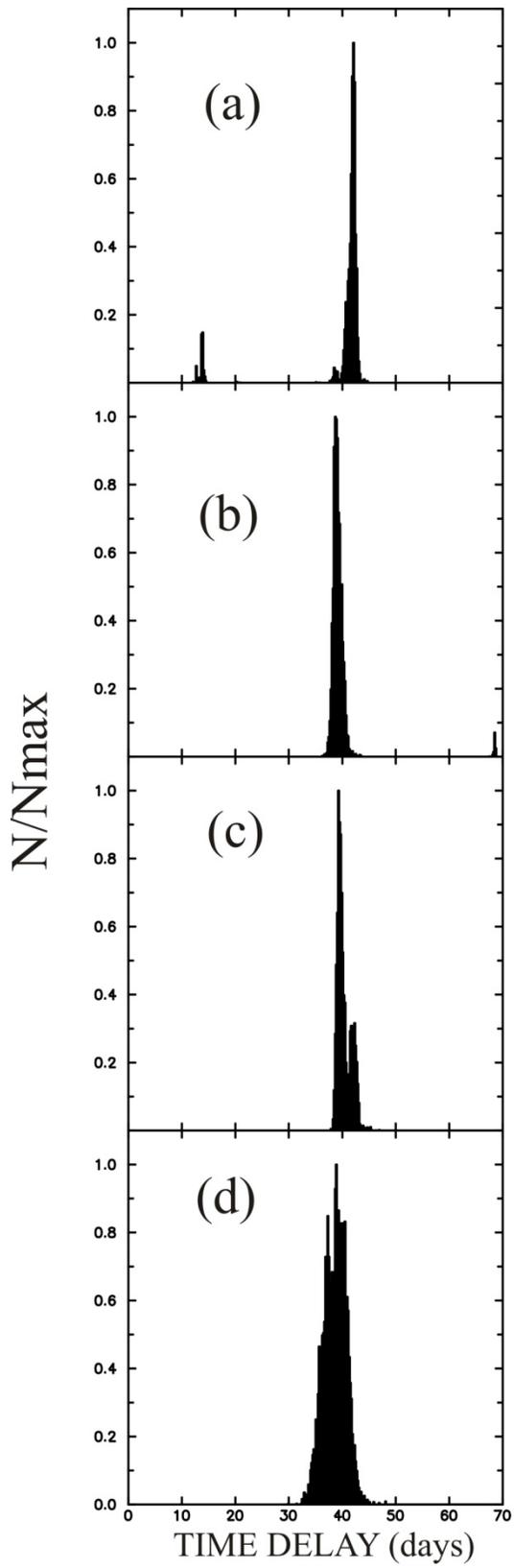

**Fig. 4.**

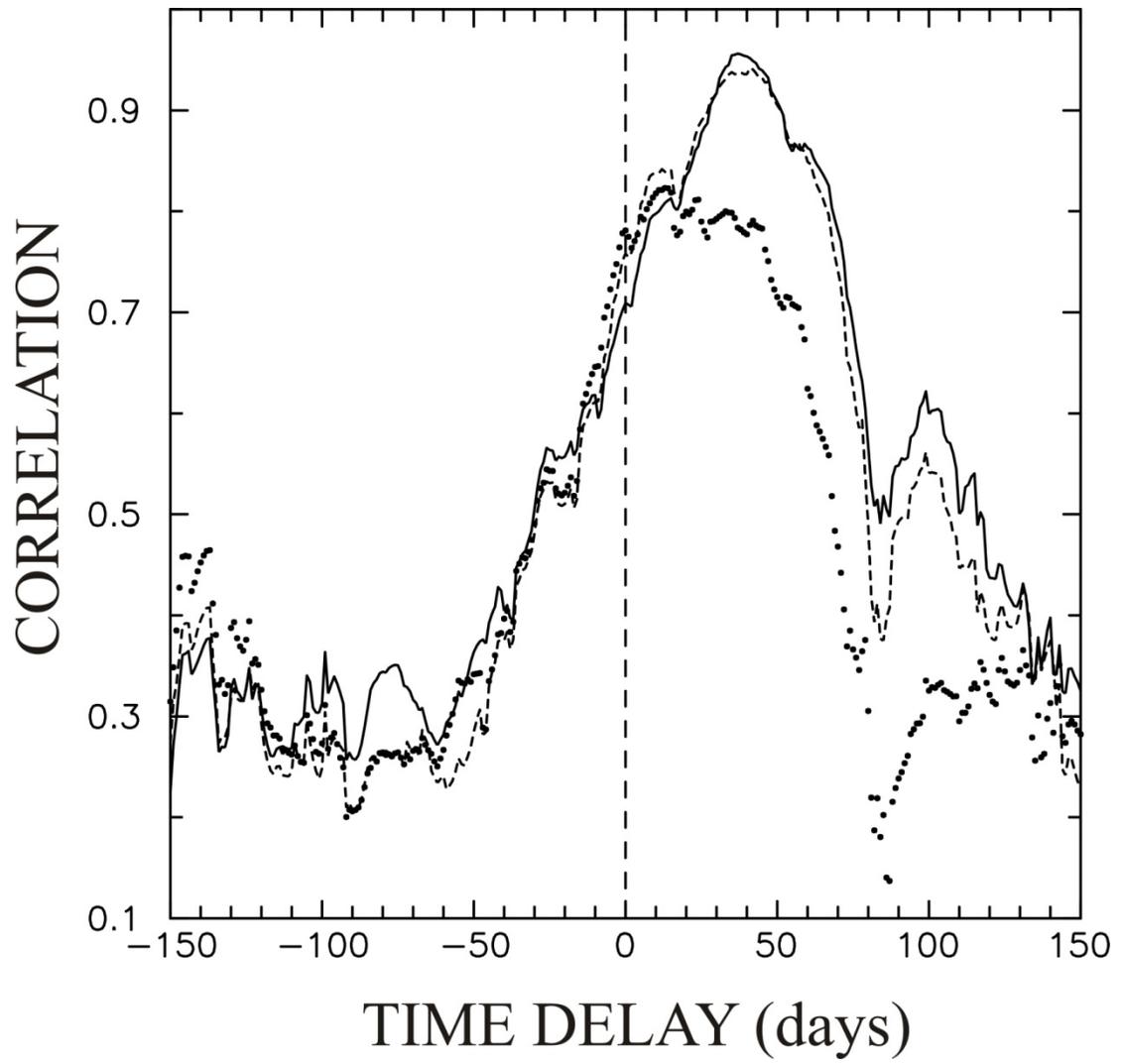

**Fig. 5.**

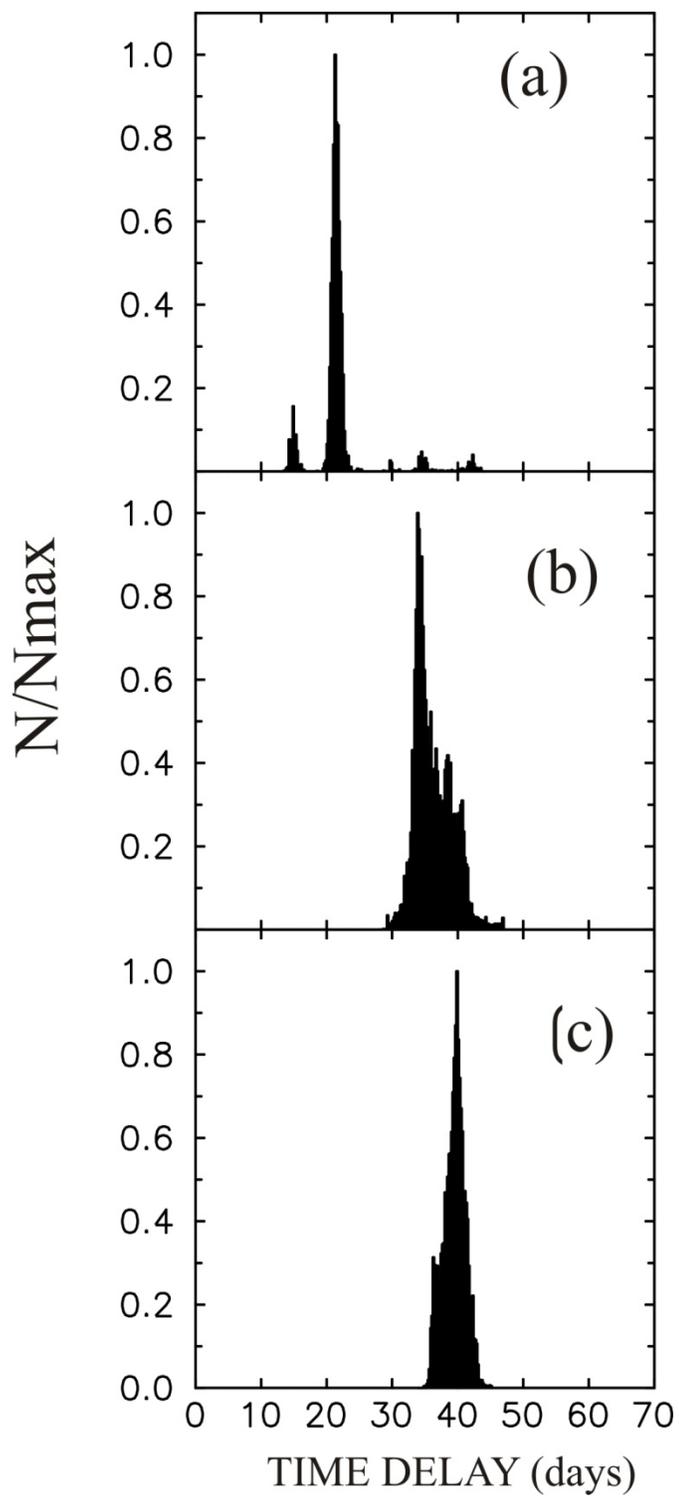

**Fig. 6.**

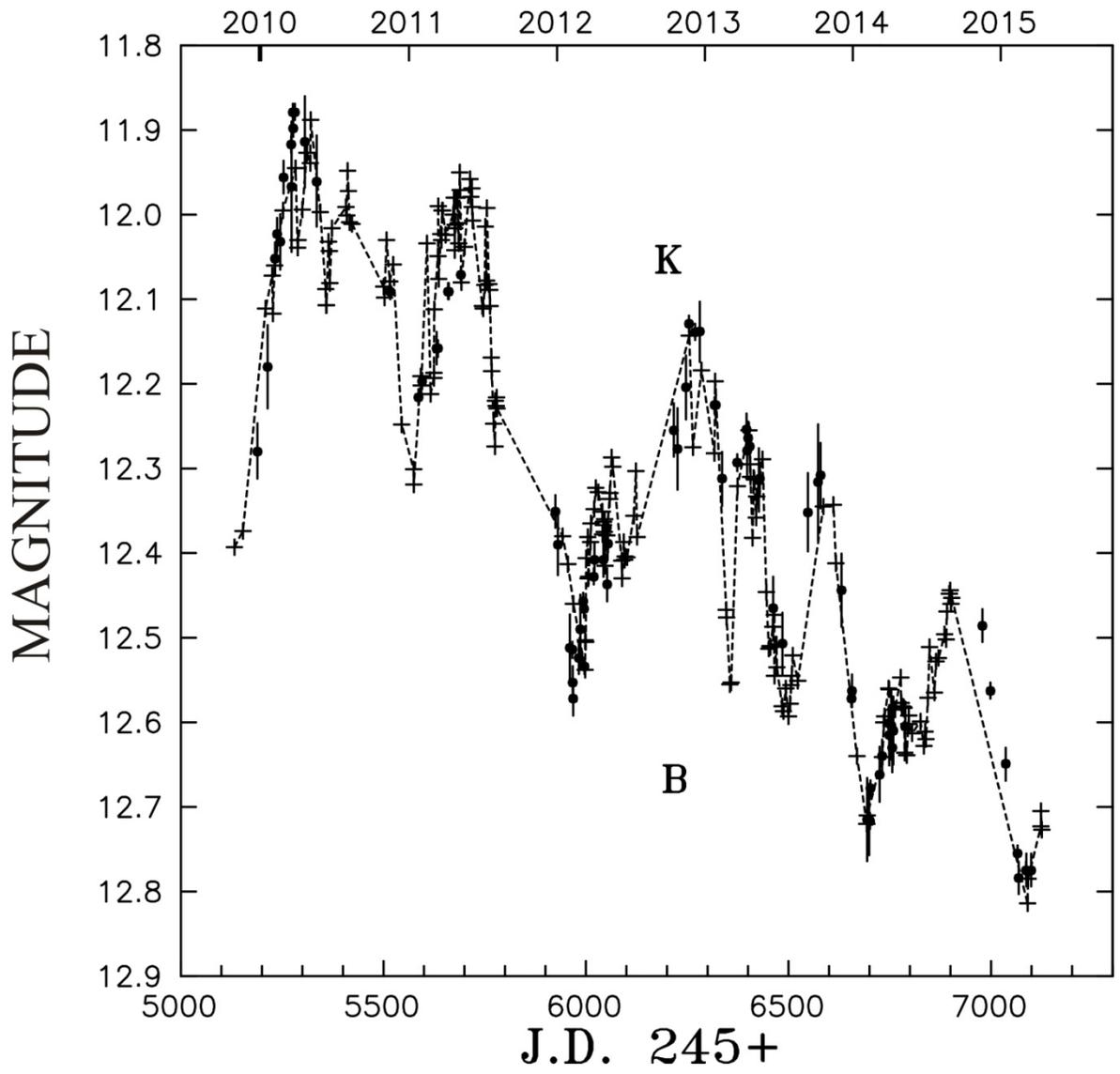

**Fig. 7.**

**CAPTURES TO THE PICTURES**

Fig. 1. Combined light curves in IR bands *JHKL* and optical *B* in 2010-2015. In IR light curves: filled circles - our data, open circles - data of Schnülle et al. (2013, 2015). In the light curve *B*: the points are our Crimean photoelectric and CCD measurements, the triangles are the reduced data of Roberts & Rumstey (2012), the crosses are the reduced dots by Guo et al. (2014), and the open circles are the reduced data of Schnülle et al. (2015).

Fig. 2. Cross-correlation functions for *K* (solid line), *H* (dashed lines), *J* (points) and *B* in the interval 20010-2015. The vertical dashed line indicates zero lag.

Fig. 3. Cross-correlation functions for *L* (dashed lines), *K* (solid line) and *B* in the interval 2010-2015. The vertical dashed line indicates zero lag.

Fig. 4. Time delay histograms obtained by the JAVELIN method for *JHKL* (respectively in Figures a, b, c, d) from *B* in the interval 2010-2015. The histograms are normalized by dividing by their maximum values.

Fig. 5. Cross-correlation functions for *K* (solid line), *H* (dashed lines), *J* (points) and *B* in the interval 2012-2014. The vertical dashed line indicates zero lag.

Fig. 6. Time delay histograms obtained by the JAVELIN method for *JHK* (respectively in Figures a, b, c) from *B* in the interval 2012-2014. The histograms are normalized by dividing by their maximum values.

Fig. 7. The light curves in *B* and *K*, where the points *K* (points) are shifted taking into account 38-day delay and are reduced to a scale *B* of values (crosses, dashed line) using a linear regression.